\documentstyle[12pt]{article}
\begin{document}
\title{{Energy Loss Effect  in  High Energy nuclear  Drell-Yan  Process}}
\author{{ChunGui Duan$^{1,5}$ LiHua Song$^{1}$ LiJuan Huo$^{2}$  GuangLie Li$^{3,4,5}$}\\
{\small 1.Department of Physics, Hebei Normal University, ShijiaZhuang ,050016,China}\\
{\small 2.physics  department, Shijiazhuang Vocational  and
Technical  institute,Shijiazhuang,050081 ,China} \\
{\small3.Institute  of  High Energy  Physics,CAS,Beijing
100039,China.} \\
{\small 4.Center  of  Theoretical Nuclear  Physics,}\\
{\small National  Laboratory  of  Heavy  Ion  Acceleration  of
Lanzhou,Lanzhou,730000,China.}\\
{\small 5.CCAST (World Laboratory).P.O.Box8730,Beijing
100080,China}
 }
\date{}
\maketitle
\baselineskip 9mm
\vskip 0.5cm
\begin{center}
\begin{minipage}{120mm}
\begin{center}
Abstract
\end{center}
The energy loss effect in  nuclear matter ,which is another
nuclear effect apart from the nuclear effect on the parton
distribution as in deep inelastic scattering process ,can be
measured best by the nuclear dependence of the high energy nuclear
Drell-Yan process.  By means of the nuclear parton distribution
studied only with lepton deep inelastic scattering experimental
data, measured Drell-Yan production cross sections for 800GeV
proton incident on a variety of nuclear targets are analyzed
within Glauber framework which takes into account energy loss of
the beam proton. It is shown that the theoretical results with
considering the energy loss effect are in good
agreement with the FNAL E866.

\end{minipage}
\end{center}

\vskip 0.5cm

{\bf I Introduction}

\hspace{1cm}Propagating of a high energy particle through a
nuclear medium is of interest in many areas of physics. High
energy proton-nucleus scattering has been studied for many decades
by both the nuclear and particle physics communities $^{[1]}$. By
means of the nuclei, we can study the space-time development of
the strong interaction during its early stages, which is
inaccessible between individual hadrons. The Drell-Yan
reaction$^{[2]}$ on nuclear targets provides, in particular, the
possibility of probing the propagation of projectile through
nuclear matter, with the produced lepton pair carrying away the
desired information on the projectile after it has travelled in
the nucleus. Only initial-state interactions are important in
Drell-Yan process since the dimuon in the final state does not
interact strongly with the partons in the nuclei. This makes
Drell-Yan scattering an ideal tool to study energy loss$^{[3]}$.

\hspace{1cm}Drell-Yan scattering is closely  related to
deep-inelastic scattering (DIS) of leptons ,but unlike DIS ,it is
directly sensitive to antiquark contributions in target parton
distributions. When DIS on nuclei occurs at $x<0.08$, where x is
the parton momentum fraction, the cross section per nucleon
decreases with increasing nucleon number A due to
shadowing$^{[4]}$. Shadowing should also occur in Drell-Yan dimuon
production at small $x_2$, the momentum fraction of the target
parton, and theoretical calculations indicate that shadowing in
the two reactions has a common origin. $^{[5]}$ The energy loss
effect is another nuclear effect apart from the nuclear effect on
the parton distribution as in DIS scattering process. Shadowing
and initial state energy loss effect  are processes that occur in
both Drell-Yan reaction and $J/\psi$ formation.Characterizing the
energy loss effect in nuclear matter should further the
understanding of $J/\psi$ production ,which is required if it is
to be used as a signal for the quark-gluon plasma in relativistic
heavy ion collisions.

\hspace{1cm}In 1999, Fermilab Experiment866(E866) $^{[6]}$
reported the precise measurement of the ratios of the Drell-Yan
cross section per nucleon for an 800GeV proton beam incident on
Be, Fe and W target at larger values of $x_1$,the momentum
fraction of the beam parton, larger values of $x_F$($\approx
x_1-x_2$),and smaller values of $x_2$  than reached by the
previous experiment, Fermilab E772$^{[7]}$.  The extended
kinematic coverage of E866 significantly increases its sensitivity
to energy loss and shadowing.

\hspace{1cm}Recently, M.B.Johnson et al.$^{[8]}$ and Francois
Arleo $^{[9]}$gave theoretical analysis of the E866 Drell-Yan
experimental data by means of different methods,
respectively.Johnson et al. examined the effect of initial state
energy loss  on the Drell-Yan  cross section ratios versus the
incident proton's momentum fraction by employing a new formulation
of the Drell-Yan  process in the rest frame of nucleus. Francois
Arleo carried out a leading-order analysis of E866  Drell-Yan data
in nuclei according to multiple scattering of a high energy parton
traversing a large nucleus ("cold " QCD matter) studied by Baier
et al.$^{[10]}$,in which the multiple soft gluon emission from the
incoming parton leads to the parton energy loss.

\hspace{1cm}Since the EMC effect was discovered, various
phenomenological models have been proposed to investigate the
nuclear effect $^{[11,12,13,14]}$ .Bickerstaff et
al.$^{[15]}$found that although most of the theoretical models
provide good explanations for the EMC effect ,they do not give a
good description of the nuclear Drell-Yan process. Most of them
overestimate the Drell-Yan different cross section ratios .In
previous paper$^{[16]}$ ,we suggested an additional nuclear effect
due to the energy loss in nuclear Drell-Yan process by means of
the Glauber model$^{[17]}$ ,which have been extensively employed
in nucleus-nucleus reaction with good fit to  related
experiment$^{[18]}$.It was found that the nuclear Drell-Yan ratio
is suppressed significantly as a consequence of continuous energy
loss of the projection proton to the target nucleon in their
successive binary nucleon-nucleon collisions. This suppression
balances the overestimate of the Drell-Yan ratio only by
consideration of the nuclear effect from the parton distribution.

\hspace{1cm}Recently,there were  two trials to obtain nuclear
parton distributions from the existing world experimental data. In
1999,K.J.Eskola et al.(EKS)$^{[19]}$ suggested a set of nuclear
parton distributions ,which are studied within a framework of the
DGLAP evolution .The measurements of $F^A_2/F^D_2$ in deep
inelastic $lA$ collisions, and Drell-Yan dilepton cross sections
measured in $pA$ collisions were used as constraints.The
kinematical ranges are $10^{-6}\leq x\leq 1$  and  $2.25GeV^2\leq
Q^2\leq10^{4}GeV^{2}$  for nuclei from deuteron to heavy ones
.With the nuclear parton distributions , the calculated results
agreed very well with the relative EMC and Fermilab E772
experimental data. In 2001,M.Hirai et al.(HKM)$^{[20]}$ proposed
two types  of nuclear parton distributions which were obtained by
quadratic and cubic type analysis. ,and determined by a $\chi^2 $
global analysis of existing experimental data on nuclear structure
functions without including the proton-nucleus Drell-Yan process
.The kinematical ranges covered $10^{-9}\leq x\leq 1$  and
 $1GeV^2\leq Q^2\leq10^{5}GeV^{2}$  for nuclei from deuteron to heavy
ones . As a result, they obtained reasonable fit to the measured
experimental data of $F_2$. In this report, by means of  the EKS
and HKM nuclear parton distribution functions ,the Drell-Yan
production cross section rations for 800Gev proton incident on a
variety of nuclear targets are analyzed by using the Glauber model
with taking into account of the energy loss of the projective
proton through nuclei.

{\bf II Method}

\hspace{1cm}According to Glauber model$^{[17]}$,  the projectile
proton scattering inelastically on nucleus (A) makes many
collisions with nucleons bound in  nuclei.The probability of
having $n$ collisions at an impact parameter $\vec{b}$ can be
expressed as
\begin{equation}
P(\vec{b},n)=\frac{A!}{n!(A-n)!}[T(\vec{b})\sigma_{in}]^{n}[1-
T(\vec{b})\sigma_{in}]^{A-n},
\end{equation}
where $\sigma_{in}$($\sim 30mb$)is non-diffractive cross section
for inelastic nucleon-nucleon collision,and $T(\vec{b})$ is the
thickness function of impact parameter $\vec{b}$. For collisions
of nucleons which are not polarized, the collisions does not
depend on the orientation of $\vec{b}$ , and $T(\vec{b})$ depends
only on the magnitude $|\vec{b}|=b$. So, we could  consider only
this case of $T(\vec{b})=T(b)$. In an nucleon-nucleus collision
without impact parameter selection, the number of nucleon-nucleon
collisions  $n$ (for $n=1$ to $A$) has a probability distribution

\begin{equation}
P(n)=\frac{\int d\vec{b}P(n,\vec{b})}{\sum\limits_{n=1}^{A}\int
d\vec{b}P(n,\vec{b})}.
\end{equation}
In following calculation , the thickness function  can be
conveniently written as$^{[18]}$
\begin{equation}
T(\vec b)=\left\{\begin{array}{cc}
\frac{1}{2\pi\beta_{A}^{2}}exp(-\vec{b}^{2}/2\beta_{A}^{2}), &  A\leq 32, \\
\frac{3}{2\pi R_{A}^{3}}\sqrt
{R^{2}_{A}-{\vec{b}}^{2}}\theta(R_{A}-|\vec{b}|), &
 A>32.
\end{array}
\right.
\end{equation}
Here $R_{A}=r_{0}A^{1/3}$ is the radius of a colliding nucleus
with $r_{0}=1.2fm$, $\theta$ is a step function , and
$\beta_{A}=0.606A^{1/3}$.

\hspace{1cm}In multiple-collision Glauber model, the basic process
is nucleon-nucleon collision for proton-nucleus Drell-Yan
process.The leading-order contribution to the Drell-Yan process is
quark-antiquark annihilation into a lepton pair. The annihilation
cross section can be obtained from the
$e^{+}e^{-}\rightarrow\mu^{+}\mu^{-}$ cross section by including
the color factor $\frac{1}{3}$ with the charge $e^{2}_{f}$ for the
quark of flavor $f$.
\begin{equation}
   \frac{d\hat{\sigma}}{dM}=\frac{8\pi\alpha^2}{9M}e^2_f\delta(\hat{s}-M^2),
\end{equation}
where $\hat{s}=x_1x_2s$,is the center of mass system (c.m.system)
energy of $q\bar{q}$ collision ,$x_1$(resp.$x_2$)is the momentum
fraction carried by the projectile (resp.target)parton, $\sqrt{s}$
is the center of mass energy of the hadronic collision, and $M$ is
the invariant mass of the produced dimuon. The hadronic Drell-Yan
differential cross section is then obtained from the convolution
of the above partonic cross section with the quark distributions
in the beam and in the target hardron:
\begin{equation}
 \frac{d^2\sigma}{dx_1dM}=K\frac{8\pi\alpha^2}{9M}\frac{1}{x_1s}
 \sum_{f}e^2_f[q^p_f(x_1)\bar{q}^A_f(x_2)
 +\bar{q}^p_f(x_1)q^A_f(x_2)],
\end{equation}
where $ K$ is the high-order QCD correction, $\alpha$ is the
fine-structure constant, and  $q^{p(A)}_{f}(x)$  and  ${\bar
q}^{p(A)}_{f}(x)$   are  the quark and anti-quark distributions in
the proton (nucleon in the nucleus A) .In addition ,one has the
kinematic relations ,
\begin{equation}
x_1x_2={\frac{M^2}{s}},\\
x_F=x_1-x_2,
\end{equation}
with the Feynman scaling variable
 \begin{equation}
x_F=\frac{2p_l}{\sqrt s},
\end{equation}
where  $p_l$ is the longitudinal momentum of the virtual photon ,

\hspace{1cm}Now let us take into account of the energy loss of the
projectile proton  moving through nuclei. In proton-nucleus
Drell-Yan collision, the incident proton passes through the
nucleus befor producing  the high $Q^2$ dimuon pair. On the one
hand ,the projectile proton may interact spectator nucleon bound
in nuclei ,in which soft (nonperturbative) minimum  bias
collisions may occur before making the final lepton pair.As a
result,the projectile proton imparts energy to the struck nucleon
and therefore must loose energy.  On the other hand, the
projectile proton, which travels through nucleus with existence of
multiple-collisions , may emit soft gluon between two bias
collisions in the nucleus ,and  must consequently experience an
energy loss. Thus the energy loss from two above  aspects in
multiple collisions can induce the decrease of  c.m.system energy
of the nucleon-nucleon collision producing  dimuon ,and affect the
measured Drell-Yan cross section. After the projectile proton has
$n$ collisions with nucleons in
 nuclei, suppose  for conveniently calculation  that the c.m.system
energy of the nucleon-nucleon collision producing  dimuon  can be
expressed as
\begin{equation}
\sqrt{s^{\prime}}=\sqrt{s}-(n-1)\triangle\sqrt{s},
\end{equation}
where $\triangle\sqrt{s}$ is the c.m.system  energy loss per
collision in the initial state . Therefore, the cross section for
Drell-Yan process can be rewritten as
\begin{equation}
\frac{d^{2}\sigma^{(n)}}{dx_{1}dM}=K\frac{8\pi\alpha^{2}}{9M}\frac{1}{x_1s^{\prime}}
\sum\limits_{f}e^{2}_{f}[q_{f}^{p}(x_{1}^{\prime})\bar{q}_{f}^{A}(x^{\prime}_{2})
+\bar{q}_{f}^{p}(x^{\prime}_{1})q_{f}^{A}(x_{2}^{\prime})].
\end {equation}
Here the rescaled quantities are defined as
\begin{equation}
x_{1,2}^{\prime}=r_{s}x_{1,2},
\end{equation}
with the c.m.system energy ratio:
\begin{equation}
r_{s}=\frac{\sqrt{s}}{\sqrt{s^{\prime}}},
\end{equation}
because of
\begin{equation}
x_{F}^{\prime}=\frac{2p_{l}}{\sqrt{s^{\prime}}}=r_{s}x_{F}.
\end{equation}
It is obvious that  $r_s$ is always greater than one. Combining
above ingredients, The measured Drell-Yan  cross section in
proton-nucleus collision experiments can be expressed as
\begin{equation}
\langle
\frac{d^{2}\sigma}{dx_{1}M}\rangle=\sum\limits_{n=1}^{A}P(n)\frac
{d^{2}\sigma^{(n)}}{dx_{1}M},
\end{equation}

{\bf III Results and Discussion}

\hspace{1cm} In order to compare with the experimental data from
E866 collaboration$^{[6]}$,we introduce the nuclear Drell-Yan
ratios as:
\begin{equation}
R_{A_{1}/A_{2}}(x_{1})=\frac{\int dM\langle
\frac{d^{2}\sigma^{p-A_{1}}}{dx_ {1} dM}\rangle}{\int dM\langle
\frac {d^{2}\sigma^{p-A_{2}}}{dx_{1}dM}\rangle}.
\end{equation}
The integral range on M is determined according to the E866
experimental kinematic region.In our theoretical analysis
,$\chi^2$ is calculated with the Drell-Yan differential cross
section rations $R_{A_1/A_2}$ as
\begin{equation}
\chi^2=\sum\limits_{j}\frac{(R^{data}_{A_1/A_2,j}-R^{theo}_{A_1/A_2,j})^2}
{(R^{err}_{A_1/A_2,j})^2},
\end{equation}
where the experimental error is given by systematic errors as
$R^{err}_{A_1/A_2,j} $, and $ R^{data}_{A_1/A_2,j}$(
$R^{theo}_{A_1/A_2,j}$) indicates the experimental data
(theoretical values ) for the ratio$R_{A_{1}/A_{2}}$.

\hspace{1cm} If the EKS$^{[19]}$ nuclear parton distribution
functions are used together with CTEQ (the coordinated Theoretical
Experimental Project on QCD )$^{[21]}$ parton distribution
functions in proton,obtained $\chi^2$ value is $\chi^2=51.4$
without energy loss  effects for the 56 total data points.  The
$\chi^2$ per degrees of freedom is given by $\chi^2/d.o.f.=0.918$.
It is shown that theoretical results agree very well with the E866
experimental data ,which results from EKS parametrization of
nuclear parton distributions studied with including the Drell-Yan
process.

\hspace{1cm}In addition,we consider also using HKM$^{[20]}$
nuclear parton distribution functions together with  MRST
$^{[22]}$parton distribution functions in proton. The calculated
results with HKM cubic type of nuclear parton distribution are
shown in Fig.1 and Fig.2. which is the Drell-Yan cross section
ratios for Fe  to Be and   W to Be as functions of $x_1$ for
various interval of $M$,respectively.The solid curves are the
ratios with only the nuclear  effect on the parton distribution as
in DIS scattering process, and the dotted curves correspond to an
energy loss effect: $\triangle\sqrt{s}= 0.18$GeV  with nuclear
effect on structure function . Obtained $\chi^2$ values are
$\chi^2=56.39$  with energy loss effect(i),$\chi^2=143.74(ii)$
without  energy loss effect,respectly,for the 56 total data
points. The $\chi^2$ per degrees of freedom are
$\chi^2/d.o.f.=1.007(i),2.567(ii)$,respectively. Employing the HKM
quadratic  type of nuclear parton distribution ,$\chi^2$ values
are $\chi^2=56.81$ with energy loss effect,$\chi^2=143.88$ without
energy loss effect,respectly.The $\chi^2$ per degrees of freedom
are  $\chi^2/d.o.f.=1.015,2.569$ , respectively. This implies that
the observed energy loss of the incident proton is
$\Delta\sqrt{s}=0.18$GeV  at $1\sigma$,which   is almost equal to
our previous results$\Delta\sqrt{s}=0.2$GeV$^{[16]}$. From
comparison with the experimental data , it is found  that our
theoretical results with energy loss effect are in good agreement
with the Fermilab E866.

\hspace{1cm}In summary, we have made a leading-order analysis of
E866 data in nuclei within the framework of the Glauber model by
taking into account of the energy loss effect of the beam
proton.In continuous multiple collisions ,the energy loss effect
in initial state causes the suppression of the proton-induced
nuclear Drell-Yan cross section . It is found that  the
theoretical results with energy loss are in good agreement with
the Fermilab E866 experiment by means of the parametrization of
nuclear parton distributions studied without nuclear Drell-Yan
process. This is another nuclear effect apart from the nuclear
effect on the parton distribution as in DIS scattering process.
The nuclear effect on structure functions  and initial state
energy loss can occur in both Drell-Yan production and $J/\psi$
formation. Hence, these researches should also further the
understanding of $J/\psi$ production. Considering the existence of
energy loss effect in Drell-Yan lepton pairs production,we think
that the determination of nuclear parton distribution functions
should not include Drell-Yan experimental data.Although there are
abundant data on electron and moun deep inelastic scattering,
valence quark distributions in the small x region and the
anti-quark distributions are difficultly determined .At this stage
,only valence quark distributions in medium x region can be
relatively well determined. It is well considered  that the
precise nuclear parton distributions must be known in order to
calculate cross sections of high energy nuclear reactions
accurately and find a signature of quark-gluon plasma in high
energy heavy-ion reactions.We suggest precise neutrino scattering
experiments ,which can provide a good  method for measuring the
$F_2(x,Q^2)$ and $xF_3(x,Q^2)$ structure functions.Using the
average of $xF_3^{\nu A}(x,Q^2)$and $xF_3^{\bar{\nu}A}(x,Q^2)$ ,
the valence quark distribution functions can be well determined.
Combining the lepton inelastic scattering data with the neutrino
scattering experiments, valence quark and anti-quark distribution
functions will be obtained in the future,which makes us good
understanding the energy loss effect in high energy nuclear
Drell-Yan collisions.

{\bf Acknowledgement:}This work is partially supported by  Major
State Basic Research Development program (under Contract No.
G20000774), by the National Natural Science Foundation of
China(No.19875024,19835010,10175074), and by Natural Science
Foundation of Hebei Province(No.103143)

\begin{center} Figure caption
\end{center}
Fig.1 The nuclear Drell-Yan cross section ratios
$R_{{A_1}/{A_2}}(x_1)$ on Fe to Be for various intervals M. Solid
curves correspond to nuclear effect on structure function  .
Dotted curves  show the combination of shadowing and energy loss
effect( $\triangle\sqrt s= 0.18$GeV) with HKM cubic type of
nuclear parton distributions .The experimental data are taken from
the E866[6].

Fig.2 The nuclear Drell-Yan cross section ratios
$R_{{A_1}/{A_2}}(x_1)$ on W to Be for various intervals M. The
comments are the same as Fig.1


\vskip 1cm
\begin{thebibliography}{s2}
\bibitem{s1}  W.Busza R Ledoux, Ann.Rev.Nucl.Part.Sci.,38(1989)119.
\bibitem{s2}  S.Drell and T.M.Yan, Phys.Rev,Lett.,25(1970)316.
\bibitem{s3}  G.T.Garvey and J.C.Peng ,Phys.Rev.Lett.,90(2003)092302.
\bibitem{s4}  M.Arneodo.et.al.(EMC), Nucl.Phys,B441(1995)3.
\bibitem{s5}  S.J.Brodsky,A.Hebecker,E.Quark.Phys.Rev.D.55.(1997)2584.
\bibitem{s6}  M.A.Vasiliev,et.al.(E866),Phys.Rev.Lett.83(1999)2304.
\bibitem{s7}  D.M.Adle et al.(E772),Phys.Rev.Lett.,64(1990)2479
\bibitem{s8}  M.B.Johnson,B.Z.Kopeliovich and I.K.Potashnikova,Phys.Rev.Lett.,86(2001)4483
\bibitem{s9}  F.Arleo,Phys.Lett.,B532(2002)231
\bibitem{s10} R.Baier,Yu.L.Dokshitzer,A.H.Mueller,S.Peigne,and D.Schiff,Nucl.Phys.,B484(1997)265
\bibitem{s11} C.A.Garacia Canal,E.M.Santangle,and H.Vucetich,Phys.Rev.Lett.,53(1984)1430
\bibitem{s12} F.E.Close,R.L.Jaffe,R.G.Roberts and G.G.Ross,Phys.Rev.,D31(1985)1004
\bibitem{s13} G.L.Li,K.F.Liu,G.E.Brown,Phys.Lett.,B213(1998)531
\bibitem{s14} Zhenmin He,Xiaoxia Yao,Chungui Duan,Guanglie Li,Eur.Phys.J.C4(1998)301
\bibitem{s15} R.P.Bickerstaff,M.C.Birse,G.A.Miller,Phys.Rev.,D33(1986)322
\bibitem{s16} Jianjun Yang and  Guanglie Li,Eur.Phys.J.C5(1998)719
\bibitem{s17} R.J.Glauber,Lecture in Theoretical Physics, edited by W.E.Brittin\\
             and L.G.Dunham(New.York,1959),vol.1,p.315.
\bibitem{s18} C.Y.Wong,Phys.Rev.D30(1984)961;C.Y.Wong,{\sl Introduction to \\
             High-Energy Heavy-Ion Collisions,}\\
             (World Scientific, Pulishing, Coe Singa Pte.Ltdl.,1994),p.249.
\bibitem{s19} K.J.Eskola,V.J.Kolinen and C.A.Salgado(EKS),Eur.Phys.J.C9(1999)61.
\bibitem{s20} M.Hirai,S.Kumano,M.Miyama(HKM) ,Phys.Rev.D64(2001)034003.
\bibitem{s21} H.L.Lai, et al.(CTEQ),Eur.Phys.J.C5(1998)461.
\bibitem{s22} A.D.Martin, R.G.Roberts, W.J.Stirling ,and R.S.Thorne(MRST),\\
              Eur.Phys.J.C4(1998)463.
\end{thebibliography}
\end{document}